\documentclass[preprint2]{aastex} 

\usepackage{amsmath}
\usepackage{natbib}

\shorttitle{ Modeling Polar Fields }
\shortauthors{Upton \&Hathaway }

\begin{document}
\title{Predicting the Sun's Polar Magnetic Fields with a Surface Flux Transport Model}
 
\author{Lisa Upton}
\affil{Department of Physics \&Astronomy, Vanderbilt University, VU Station B 1807, Nashville, TN
37235 USA}
\affil{Center for Space Physics and Aeronomy Research, The University of Alabama in Huntsville,
Huntsville, AL 35899 USA}
\email{lisa.a.upton@vanderbilt.edu}
\email{lar0009@uah.edu}

\author{David H. Hathaway}
\affil{NASA Marshall Space Flight Center, Huntsville, AL 35812 USA}
\email{david.hathaway@nasa.gov}

\begin{abstract}
The Sun's polar magnetic fields are directly related to solar cycle variability. The strength of the polar fields at the start (minimum) of a cycle determine the subsequent amplitude of that cycle. In addition, the polar field reversals at cycle maximum alter the propagation of galactic  cosmic rays throughout the heliosphere in fundamental ways. We describe a surface magnetic flux transport model that advects the magnetic flux emerging in active regions (sunspots) using detailed observations of the near-surface flows that transport the magnetic elements. These flows include the axisymmetric differential rotation and meridional flow and the non-axisymmetric cellular convective flows (supergranules) all of which vary in time in the model as indicated by direct observations. We use this model with data assimilated from full-disk magnetograms to produce full surface maps of the Sun's magnetic field at 15-minute intervals from 1996 May to 2013 July  (all of sunspot cycle 23 and the rise to maximum of cycle 24). We tested the predictability of this model using these maps as initial conditions, but with daily sunspot area data used to give the sources of new magnetic flux. We find that the strength of the polar fields at cycle minimum and the polar field reversals at cycle maximum can be reliably predicted up to three years in advance. We include a prediction for the cycle 24 polar field reversal. 

\end{abstract}


\section{INTRODUCTION}

Obtaining a complete understanding of solar cycle variability is one of the oldest and most significant problems in solar physics. \citet{Babcock59} reported the first observation of a reversal in the Sun's dipolar magnetic fields, noting that this reversal occurred near the time of solar maximum. Shortly thereafter, he linked solar cycle variability to magnetism on the Sun by proposing a solar dynamo model \citep{Babcock61}. While Babcock's model is widely accepted as the underlying mechanism behind the solar cycle, the finer details are still not well understood. 

Two points in the evolution of the polar fields stand out as being the most significant: the reversal of the polar fields and the polar fields at solar minimum. The reversal of the polar fields marks the time of solar cycle maximum, i.e. when solar activity begins to wane. Furthermore, the reversal is important to cosmic ray observations. The polarity of the solar dipole changes the manner in which the positively charged cosmic rays propagate through the heliosphere \citep{FerreiraPotgieter04}. This results in cosmic ray flux with a flat peak when the Sun's magnetic dipole is positive and a sharp peak when the Sun's magnetic dipole is negative. On the other hand, the polar fields at solar minimum are thought to be the seeds to the next solar cycle. Indeed, observations have shown that the strengths of the polar fields at solar minimum are a good indicator of the strength of the next cycle \citep{Svalgaard_etal05,MuozJaramillo_etal12,SvalgaardKamide13}. Interestingly, the polar fields leading up to the Cycle 23/24 minimum were about half as strong as observed for the previous two cycles \citep{Svalgaard_etal05}. This was followed by an extended Cycle 23/24 minimum and what is proving to be the weakest solar cycle in over a hundred years. This has caused speculation that the Sun may be entering another Maunder Minimum. With such unusual solar conditions there is increasing motivation to determine exactly how magnetic flux is transported to the poles and how the polar fields are modulated.

The dynamo model of \citet{Babcock61} can be broken into two fundamental processes: 1. The conversion of the Sun's poloidal field at minimum into toroidal field of sunspots, and 2. The conversion of the toroidal magnetic field into reversed poloidal magnetic field. Surface flux transport models \citep{DeVore_etal84, Wang_etal89, vanBallegooijen_etal98, SchrijverTitle01} focus on the latter process. To begin this latter process, magnetic flux is taken to emerge in active regions with a characteristic tilt, i.e. Joy's Law tilt \citep{Hale_eta19, Howard1991}, and is then shredded off into the surrounding plasma. The lower latitude leading polarity flux cancels across the equator and the surface flows transport the higher latitude following polarity flux to the poles. The following polarity cancels with the original poloidal fields and creates new poloidal fields with opposite polarity. 

Previous surface flux transport models have used meridional flow profiles that worked best with the model. 
These flow profiles were constant in time and typically stopped completely before reaching $75^{\circ}$. However, recent observations have shown that these meridional profiles are not realistic. The meridional circulation has been found to vary considerably over the solar cycle and from one cycle to the next \citep{HathawayRightmire10, BasuAntia10, GonzalezHernandez_etal10}. Furthermore, the meridional flow can extend all the way to the poles \citep{HathawayRightmire11, RightmireUpton_etal12}. Idealy, a transport model should be able to reproduce the magnetic field evolution at the surface by incorporating the observed flows.

Previous surface flux transport models employed a diffusive term to simulate effects of convective motions. In section 2, we introduce a purely advective surface flux transport model. This model is used to investigate the evolution of Sun's polar magnetic fields. In section 3, we assimilate magnetic field data from full disk magnetograms into the surface flux transport model. This ensures that it accurately represents the magnetic fields on the entire surface of the Sun. This baseline is used in section 4 to illustrate the difference in the timing of the polar field reversals based on four different definitions of polar fields. Advantages and disadvantages of each these definitions are discussed. 

In Section 5, we outline the steps needed to modify the surface flux transport model for predictive purposes, incorporating data from active region databases to simulate active region emergence. We demonstrate this technique using Solar Cycle 23 active region data to reproduce the axial magnetic dipole moment leading up to the Cycle 23/24 minimum. In Section 6, we investigate the accuracy of predictions made with this model using proxy data for the active region. We model the flux transport for two phases of the solar cycle: leading up to the Solar Cycle 23/24 minimum and the reversal of the Sun's axial dipole moment during Solar Cycle 23 maximum. The predicted axial dipole moment leading up to solar minimum is compared to the baseline dipole moment to determine its accuracy in amplitude. The predicted dipole reversal is compared to the baseline reversal to determine the accuracy in the timing of these predictions. Finally, in Section 7 we use the modified surface flux transport model to examine the status of and make predictions about the current (Solar Cycle 24) polar field reversal as measured by the axial dipole.

\section{SURFACE FLUX TRANSPORT MODEL}   

We have created a surface flux transport model to simulate the dynamics of magnetic fields over the entire surface of the Sun. The basis of this flux transport model is the advection equation:
\begin{align}
\frac{\partial B_{r}}{\partial t} +  \nabla \cdot (uB_{r}) &= S(\theta,\phi,t) 
\end{align}
where $B_{r}$ is the radial magnetic flux, $u$ is the horizontal velocity vector (which includes the observed axisymmetric flows and the convective flows), and S is a (magnetic) source term as a function of latitude, longitude, and time.

This purely advective model is supported by both theory and observation. The Sun's magnetic field elements are carried to the boundaries of the convective structures (granules and supergranules) by flows within those convective structures. The motions of those magnetic elements are faithful representations of the plasma flow itself. These weak magnetic elements are transported like passive scalars (corks). This has been found in numerous numerical simulations of magneto-convection (c.f. \citep{Vogler_etal05}) and is born out in high time- and space-resolution observations of the Sun \citep{Simon_etal88, Roudier_etal09}. 

The axisymmetric flows (meridional flow and differential rotation) have been measured for each Carrington rotation by using feature tracking on MDI and HMI magnetograms \citep{HathawayRightmire10, HathawayRightmire11, RightmireUpton_etal12}. These axisymmetric flow profiles were fit with polynomials and the polynomial coefficients were smoothed using a tapered Gaussian with a full width at half maximum of 13 rotations. These smoothed coefficients were used to update the axisymmetric flow component of the vector velocities for each rotation, thereby including the solar cycle variations inherent in these flows.

The convective flows, i.e. supergranular flows, were modeled explicitly by using vector spherical harmonics, as described by \cite{Hathaway_etal10}. A spectrum of spherical harmonics was used to create convection cells that reproduce the observed spectral characteristics. The spectral coefficients were evolved at each time step to give the cells finite lifetimes and the observed differential rotation and meridional flow. These convection cells have lifetimes that are proportional to their size, e.g. granules with velocities of 3000 m s$^{-1}$, diameters of 1 Mm, and lifetimes of $\sim$10 minutes and supergranules  with velocities of 300 m s$^{-1}$, diameters of 30 Mm, and lifetimes of $\sim$1 day. These convective cells are advected by the axisymmetric flows given by the smoothed polynomial coefficients. The vector velocities were created for the full Sun with 1024 pixels in longitude and 512 pixels in latitude at 15 minute time steps.

Outside of active regions, the magnetic fields are weak and the plasma beta is high. The magnetic pressure of these weak fields is dominated by the kinematic pressure, and these weak fields are carried by the plasma flows. Inside active regions, the plasma beta is high and the flows are modified by the magnetic field. To account for this, we reduced the supergranule flow velocities where the magnetic field was strong. 

The surface flux transport advection equation was solved with explicit time differencing to produce magnetic flux maps of the entire Sun with a cadence of 15 minutes. (These maps are referred to as synchronic maps since they represent the Sun's magnetic field at a moment in time.) To stabilize the numerical integrations and avoid ringing artifacts due to the Gibbs phenomena, we added a diffusion term so that:
\begin{align}
\frac{\partial B_{r}}{\partial t} +  \nabla \cdot (uB_{r}) &= S(\theta,\phi,t) + \eta \nabla^{2}B_{r} 
\end{align}
where $\eta$ is the diffusivity. We note that this diffusivity term was strictly for numerical stability. The addition of this term had little effect on the flux transport. The convective motions of the supergranular cells gave detailed random walks for the magnetic elements in this model.

\section{BASELINE MODEL: DATA ASSIMILATION}    

We used the flux transport model to create a baseline dataset. For this baseline, we assimilated full disk magnetograms to provide the closest contact with observations. The process of data assimilation periodically updates and corrects for any differences between data and model. Regions where data were recently assimilated are nearly identical to the observations. This baseline was used to examine the different methods for characterizing the polar fields and also served as a metric for evaluating the prediction results in Section 6.

Magnetograms were obtained from the Michelson Doppler Imager (MDI) \citep{Scherrer_etal95} onboard  
\emph{Solar and Heliospheric Observatory} (SOHO) and the Helioseismic Magnetic Imager (HMI) onboard the \emph{Solar Dynamics Observatory} \citep{Scherrer_etal12}. MDI magnetograms were assimilated from 1996 May to 2010 May with a cadence of every 96 minutes (excluding the time period in late 1998 and early 1999 when MDI data was unavailable or unreliable). From 2010 May to 2013 July the HMI magnetograms were assimilated hourly. The flux in each pixel of the magnetograms was divided by the cosine of the angle from disk center in order to best approximate the assumed radial magnetic field. 

The data assimilation process merged magnetogram observations with forecasts made by the surface flux transport model. This was done by assigning weights to both the observed data and the data forecasted by the model (shown in Figure 1). The observed magnetic fields have signal to noise ratios that degrade away from disk center so the weights for the observed data fell off as a function of center to limb distance. The weights for the forecasted data were created by adding the newly observed weights to the model weights from the previous time step and then multiplying by a latitude dependent exponential decay function. This exponential decay function was designed to account for the drift between observations and model for places and times that observations are unavailable. The weights decay by a factor of 1/e in $\sim$1 week at the rapidly evolving equator, but more slowly (up to several months) at the poles. A new map was created by adding the forecasted data multiplied by its weights to the observed data multiplied by its weights and then by dividing by the sum of the two weights.

\begin{figure}[ht!]       
\plotone{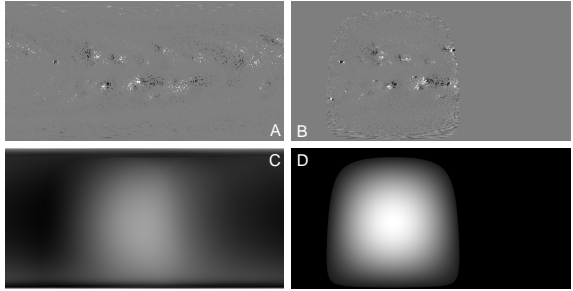}
\caption{ Data Assimilation. A) The data forecasted by the flux transport model. B) The data observed with a magnetograph. C) The weights for the simulated data. D) The weights for the observed data.
}
\end{figure}

Full Sun synchronic maps were retained at 8 hour intervals (times of 0, 8, and 16 hours), from 1996 May to 2013 July. A magnetic butterfly diagram was constructed by averaging $B_{r}$ over longitude for all of the synchronic maps in each solar rotation. This butterfly diagram, shown in Figure 2, illustrates several important details. As expected, this baseline magnetic butterfly diagram is nearly indistinguishable from a butterfly diagram constructed directly from observations. In particular, an annual signal in the polar field strength is seen at high latitudes. This annual signal has been a characteristic feature of MDI, Mount Wilson Observatory (MWO) and SOLIS datasets albeit with differences depending on the instrument and spectral line used. There have been attempts \citep{UlrichTran13,Jin_etal12} to explain the origin of this annual signal in terms of a systematic tilt of the fields, but so far there is no consensus. Perhaps one of the most telling aspects of this annual signal is that it is either not present or too weak to be seen in the HMI data. This suggests that this annual signal could be due to changes in spatial resolution, noise levels at the poles, or possibly errors (at high latitudes) in the calculation of field strength using different spectral lines. 

The baseline butterfly diagram also illustrates some important details about flux transport. Figure 2 shows that it takes $\sim$1-2 years for active region flux to be transported to the poles from the active latitudes. This suggests that a flux transport model should be able to reproduce the polar fields at least this far in advance. Furthermore, our flux transport continued during the ``SOHO summer Vacation'' from 1998 June through 1999 February, i.e. a period when no data assimilation was occurring. This resulted in the poleward transport of leading polarity flux from the lower latitudes, thus it is essential that new active region sources continue to be added. If the active region emergence is prematurely cut-off, excess leading polarity (that  would have been canceled by the new emerging flux) remains and is transported to the poles along with (or just after) the following polarity flux. This has the effect of slowing down the reversal (or depending on the timing, slowing the subsequent buildup of new polarity). If enough excess leading polarity is transported to the poles, then a relapse in the polar field reversal may also be observed. In this case, the assimilation corrected for these problems once it was re-initiated in 1999 February.

\begin{figure}[ht!]  
\plotone{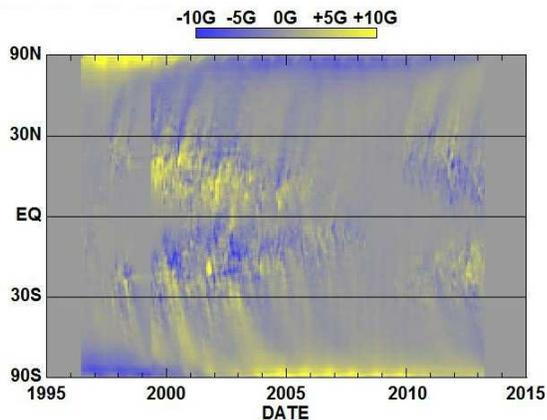}
\caption{ Baseline Magnetic Butterfly Diagram. Yellow (positive) and blue (negative) streamers show that it takes $\sim$1-2 years for active region flux to reach the poles. The ``SOHO summer Vacation'' from June 1998 through February 1999 illustrates the importance of continued active region emergence.  
}
\end{figure}

\section{POLAR FIELDS}    

Magnetic maps of the entire Sun provide the ability to change the angle from which the Sun is viewed, e.g. looking directly down on the poles as shown in Figure 3. Seeing the Sun from above the poles is vital to furthering our understanding of the evolution of polar regions and their impact on the solar cycle  \citep{Shiota_etal12, MuozJaramillo_etal13}. By watching the flux transport from this angle it is clear that the residual active region flux at high latitudes is substantially sheared by differential rotation. The combined effect of the differential rotation shearing and the meridional flow driving the flux poleward causes the residual flux to spiral into the pole. The polarity of this residual flux is typically opposite in sign to the polarity of the pole at the beginning of the solar cycle. As this (typically) opposite polarity flux reaches the poles it cancels with the original polar fields until it disappeared completely and the new (opposite polarity) polar field begins to build.        

\begin{figure}[ht!]  
\centerline{\includegraphics[width=\columnwidth]{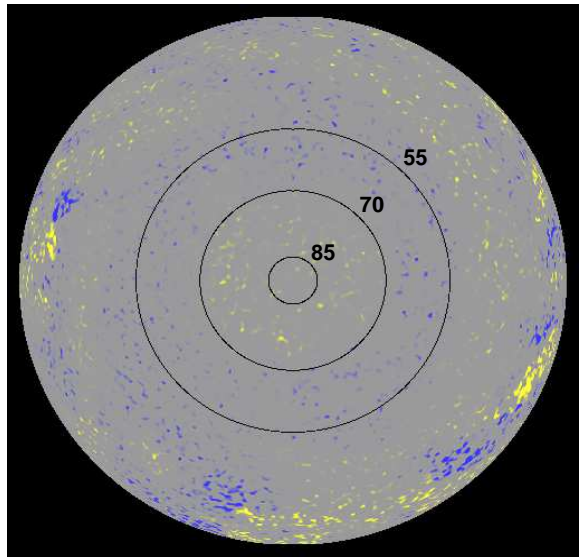}}
\caption{
North Polar View in 2001 April. Synchronic magnetic maps allow the Sun to be seen from the perspective of looking directly down on the poles. The $55^{\circ}$, $70^{\circ}$, and $85^{\circ}$ latitude lines used in definitions of polar fields are marked for reference. }
\end{figure}

The polar fields are often characterized by averaging the flux density over a polar region of the Sun. However, what area is considered a polar region is rather arbitrary. For the Wilcox Solar Observatory, the polar field strengths are defined using the line of sight fields between $55^{\circ}$ and the poles. This range is established by the resolution of the instrument. With the advancements in the spatial resolution of more modern instruments, recent polar field measurements have become more restrictive. \citet{deToma2011} measured the polar fields using the radial fields between $60^{\circ}$ and $80^{\circ}$ latitude. \citet{MuozJaramillo_etal12} obtained polar field strengths by averaging the line-of-sight fields poleward of $70^{\circ}$. Alternatively, the polar fields can be defined by the axial component of the Sun's magnetic dipole \citep{Svalgaard_etal05}.

Magnetic maps of the entire Sun also provided the benefit of being able to calculate the polar field strengths using all longitudes and latitudes extending all the way to the poles. This was done using three different definitions of polar area (above $55^{\circ}$, above $70^{\circ}$, and above $85^{\circ}$ latitude). These latitudes are indicated by the circular  black lines in Figure 3. The magnetic maps made by assimilating the MDI magnetograms produced an annual signal in the polar field strength. Fortunately, there is almost a full year of overlap (2010 April to 2011 March) in the observations of MDI and HMI. We have taken advantage of this overlap in observations to calibrate our MDI based polar field measurements. First the MDI based polar field measurements were smoothed using a tapered Gaussian with a full width at half maximum of 13 rotations. The smoothed MDI based polar field measurements were then compared to the HMI based polar field measurements. It was found that the two measurements agreed when 0.5 Gauss was uniformly subtracted from the MDI based polar field measurements. The 13 rotation tapered Gaussian smoothing and 0.5 G offset were then applied to all of the MDI data.

The corrected polar field strengths during Solar Cycle 23 maximum are shown in Figure 4 (top plot). For all three definition of polar area, the timing of the North and South reversals are well synchronized (i.e. they occur within a couple months of each other). However, the timing of the reversal varies by $\sim$1 year depending on which definition of polar area was used. For $55^{\circ}$ and above, the reversal comes at the end of 2000, the $70^{\circ}$ reversal occurs in mid-2001, and the $85^{\circ}$ reversal does not occur until the end of 2001. These results demonstrate that measuring the polar field strength over a polar area is both arbitrary (because there is no formal standard as to what polar area should be used) and ambiguous (varies by as much as a year depending on what polar area is used). 

\begin{figure}[ht!]  
\centerline{\includegraphics[width=\columnwidth]{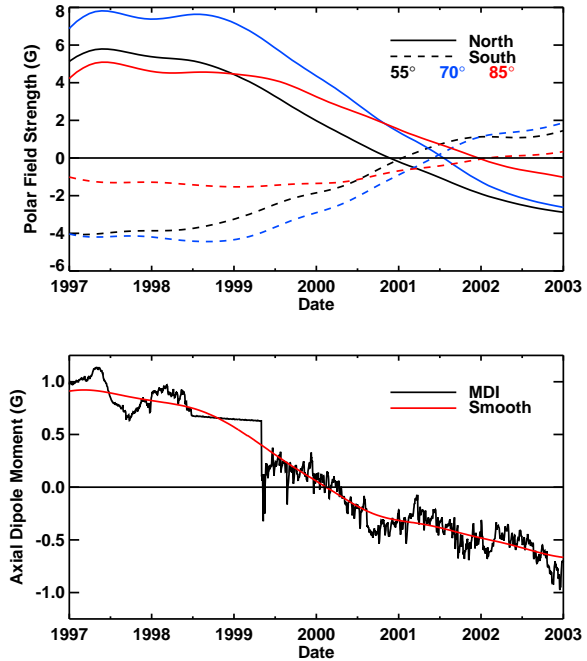}}
\caption{
Polar Field Reversals. The corrected North (solid) and South (dashed) polar field strength reversals (top) are shown for 3 different definitions of polar area: $> 55^{\circ}$ in black, $> 70^{\circ}$ in blue, and $> 85^{\circ}$ in red. The timing of the reversal depends greatly on which polar area is used, with $\sim$1 year between the $55^{\circ}$ reversal and the $85^{\circ}$ reversal. This behavior is consistent with the notion of new polarity flux spiraling in and canceling the old polarity flux residing in the polar cap. The reversal of the axial dipole moment (bottom) occurs early in 2000. The raw data (black) is contaminated by an annual signal in the MDI data. The smoothed axial dipole moment is shown in red.}
\end{figure}

The synchronic maps can be used to calculate the axial magnetic dipole moment $B_{p}$ (shown in bottom panel of figure 4), where:
\begin{align}
B_{p}  &= \int_{0}^{2\pi} \int_{0}^{\pi} B_{r}(\theta,\phi) Y_{1}^{0}(\theta,\phi) \sin\theta d\theta d\phi
\end{align}
Not surprisingly, the synchronic maps made by assimilating MDI data produced an annual signal in the magnetic dipole moment measurements as well. This was removed by smoothing with the tapered Gaussian with a full width at half maximum of 1 year (the red line in Figure 4). The axial magnetic dipole moment reverses sign in early 2000, almost precisely the time of the solar cycle 23 maximum. No data was assimilated during late 1998 and early 1999. During this time period, the axial dipole moment appears to decay very slightly and is followed by a sudden jump when data assimilation is re-initiated. For this time period in particular (and a few months afterwards) the smoothed dipole moment is a better measure of the axial dipole moment on the Sun.

The axial dipole moment appears to be a better metric for analyzing the relationship between the polar fields and the solar activity cycle. Firstly, the axial dipole moment depends on the magnetic field over the entire Sun rather some arbitrary polar area, and therefore it is less ambiguous. Secondly, the polar field strengths can become asymmetric if active region emergence is asymmetric. This is certainly an interesting and important aspect of the solar dynamo, however it is uncertain what role, if any, these asymmetries play in modulating the solar activity cycle. These hemispheric asymmetries are short lived (usually less than a year or two) and self-correcting \citep{NortonGallagher10}. As the axial dipole moment reflects the magnetic state of the Sun as a whole, it is not as sensitive to these hemispheric differences. Furthermore, the timing of the axial dipole moment reversals appears to be better correlated to the timing of the solar cycle maximum. The smoothed Wilcox Solar Observatory axial dipole moment reversed in 1979 November, 1989 December, and 1999 October. A 13 month running mean of the International Sunspot Number shows these reversals nearly coincide with solar maximum: 1979 December, 1989 July, and 2000 April. In the case of the later two, the axial dipole moment reversals actually precede solar cycle maximum by a few months, further indication that the dipole moment is a key measure of the dynamo process.

\section{PREDICTIVE MODEL: ACTIVE REGION SOURCES}      

The surface flux transport model presented here needs to be modified in order to use it for predictive purposes. Detailed predictions of the emergence of active region flux are not possible. However, reliable predictions of the number of active regions and the latitudes at which they emerge are available once a cycle is underway \citep{Hathaway10}. These predictions (or active region data from similar sunspot cycles) can be used to provide the active region sources for the flux transport model. In addition, the synchronic maps used as initial conditions need adjustments.

An initial synchronic map is needed to begin a prediction. Synchronic maps generated using the MDI data had the annual signal described above. The nature of this annual signal is still not fully understood so properly correcting the full disk magnetograms is not feasible at this time. This  flux error would propagate through the simulation and cause errors in the polar field strength measurements. The annual signal can, however, be removed from the axial dipole moment component. This is done by measuring the axial dipole moment present in each synchronic map during the MDI time period (1996 May to 2010 May), smoothing it with the 1 year tapered Gaussian, and producing a new set of maps using the smoothed axial dipole moment. The annual signal did not appear in the HMI data, and so these steps are not be necessary when a synchronic map generated from HMI data is used to initialize the simulation.

Active region emergence was simulated by adding bipolar Gaussian spot pairs in the location of the active regions. The Royal Greenwich Observatory (RGO) and the National Oceanic and Atmospheric Administration (NOAA) sunspot records provide information about the size, and location of nearly all the sunspot groups that have been observed since 1874 (solar cycles 12-24). These databases were used to characterize the active regions in terms of flux, Joy's Law tilt, and longitudinal separation. The flux was calculated as a function of reported area using the relationship described by \citet{Dikpati_etal06}:
\begin{align}
\Phi(A)  &= 7.0\times10^{19}A 
\end{align}
where $\Phi(A)$ is the magnetic flux in Maxwells and $A$ is the total sunspot area in units of micro Hemispheres ($3\times10^{16}$cm$^{2}$). The tilt was given by the average Joy's Law tilt, i.e. the angle between the bipolar spots (with respect to lines of latitude) is equal to one half of the latitude. While the NOAA sunspot record (1974 to present) includes both the sunspot area and longitudinal extent, the RGO data only include the sunspot area. Using the NOAA data, we have found a relationship between the area of the sunspot group and the longitudinal extent (shown in Figure 5): 
\begin{align}
\Delta\phi(A)  &= A\frac{17}{2000} + 7\tanh\frac{A}{70}  
\end{align}
where $\Delta\phi$ is the longitudinal extent in degrees and $A$ is the group sunspot area in micro Hemispheres. This equation was used to set the longitudinal separation of the bipolar spots added to the simulation from the RGO database.

\begin{figure}[ht!]  
\plotone{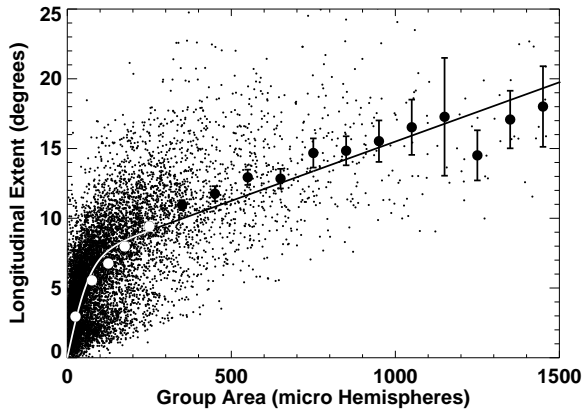}
\caption{Longitudinal Extent of Sunspot Groups. NOAA data from 1995 to 2013 was used in this plot of the longitudinal extent of individual sunspot groups as a function of group area. The large dots show the averaged binned data with 2 sigma errors. Equation 5 is shown as the solid line.  
}
\end{figure}

Lastly, instead of using the measured flows for each rotation in the flux transport, the average axisymmetric flows were used to create the vector velocities in these prediction simulations. Alternatively, one could generate meridional flow profiles that have the observed systematic solar cycle variations \citep{BasuAntia03, HathawayRightmire10}. Future work will investigate the importance of these systematic meridional flow variations in this flux transport model.  

To demonstrate the viability of this predictive flux transport model, we simulated the magnetic field evolution for the 3 years leading up to the Solar Cycle 23/24 minimum using the active regions from Solar Cycle 23. The simulation is repeated 5 times, using different realizations of the supergranular flows. Statistically these realizations all had cellular flows with the same characteristic sizes and lifetimes, but the details of the individual cells was changed (e.g. their locations relative to active region flux concentrations).

All 5 realizations of the axial dipole moment evolution are shown in Figure 6. For comparison, the unsmoothed baseline axial dipole moment is shown by the dashed black line. All of the realizations are in good agreement, showing a dipole moment that coincides almost precisely with the baseline dipole moment. The increase in the spread of the measurements over time highlights the stochastic nature and important role that supergranules play in the transport of flux. The random details of individual cells can produce variations in the dipole moment on time scales of years, but this variation is significantly smaller than the variation due to the annual signal carried over from the MDI data. Despite these stochastic variations, the flux transport demonstrates its functionality and potential for predicting the polar fields 3 years (and perhaps longer) in advance of solar cycle minimum.

\begin{figure}[ht!]    
\centerline{\includegraphics[width=\columnwidth]{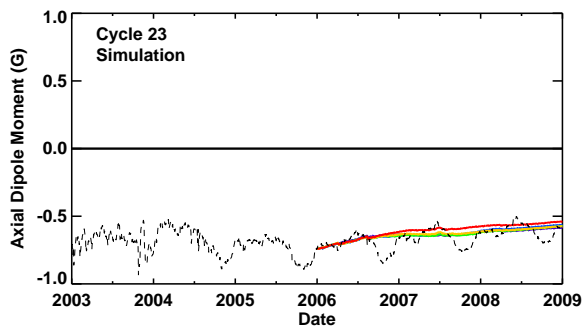}}
\caption{
Predictions of Cycle 23/24 axial dipole moment approaching minimum using Cycle 23 active regions starting $\sim$3 years ahead of the observed minimum. The 5 different supergranule realizations are represented by the solid lines shown in color. For reference the unsmoothed MDI axial dipole moment is shown with a black dashed line. }
\end{figure}

\section{PREDICTION TESTS - CYCLE 23 USING CYCLE 17 ACTIVE REGIONS}       

We tested the predictive abilities of this flux transport model by attempting to reproduce the axial dipole moments of Solar Cycle 23 using proxy data for active region sources. Solar Cycle 17 most closely matched the amplitude and shape of Solar Cycle 23 (shown in Figure 7) and  was used as a proxy to for Solar Cycle 23 active region emergence. We investigated two primary points of interest during the solar cycle: Solar Cycle 23/24 minimum (the end of 2008) and the reversal of the polar fields during Solar Cycle 23 maximum (spring of 2000). In both cases the model started with a lead time of $\sim$3 years and 5 different realizations of the convective motions were used. The simulation of the polar field reversal ran until the end of 2002, to ensure that the reversal was fully captured.  

\begin{figure}[ht!]    
\centerline{\includegraphics[width=\columnwidth]{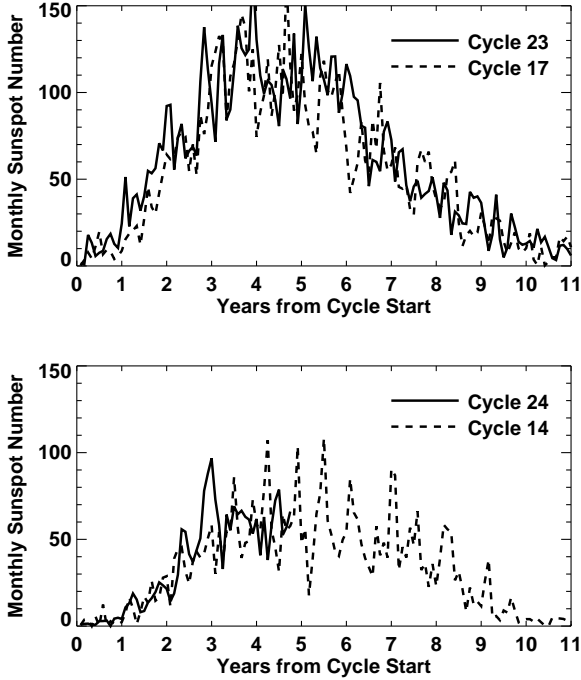}}
\caption{
 Solar Cycle Proxies. Active region sources are simulated by using prior solar cycles as proxies for the modeled cycles. Solar Cycle 17 is chosen as a proxy for Solar Cycle 23 (top). Solar Cycle 14 is chosen as a proxy for Solar Cycle 24 (bottom).}
\end{figure}

The predictions of the approach to Solar Cycle 23/24 minimum using Cycle 17 active region data (SC23AR17) is shown in the top panel of Figure 8. For comparison, the baseline axial dipole moment is shown by the dashed black line. The SC23AR17 prediction is fully consistent with the baseline. For the first two years, the axial dipole moments are nearly identical to the axial dipole moments that were simulated using the Cycle 23 active regions (SC23AR23, Figure 6). For the last year, the SC23AR17 prediction begins to diverge somewhat from the SC23AR23 simulation. This divergence is small in comparison to the annual signal variation seen in the baseline.

\begin{figure}[ht!]    
\centerline{\includegraphics[width=\columnwidth]{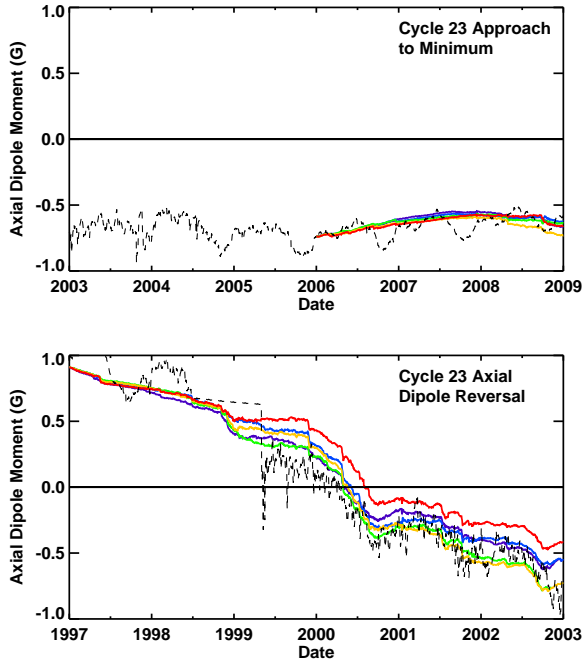}}
\caption{
Predictions of Cycle 23 with Cycle 17 active regions. Axial dipole moment predictions of Cycle 23/24 Minimum made with Cycle 17 active regions and a start time of $\sim$3 years (top). Axial dipole moment predictions of Cycle 23 polar fields reversal made with Cycle 17 active regions and a start time of $\sim$3 years (bottom). In both cases, the 5 different realizations are represented by the solid lines shown in color. For reference the unsmoothed MDI axial dipole moment is shown with a black dashed line. }
\end{figure}

The predictions of the SC23AR17 dipole moment reversal is shown in the bottom panel of Figure 8. All of the realizations (made with a lead time of $\sim$3 years ahead) predicted the timing of the reversal to within four months of the baseline axial dipole moment reversal. Four of the realizations predict the timing of the reversal almost precisely (to within a month). The fifth realization places the reversal about four months late. Surprisingly, the amplitude of the dipole moment (in 4 of the 5 realizations) stays in remarkably good agreement with the baseline through to the end of the prediction (some six years after the prediction start time).

Comparison of the model predictions during the two different phases of the solar cycle shows that the spread of the measurements (due to the stochastic nature of supergranules) is more pronounced in the prediction for the dipole moment reversal (i.e. solar maximum) than the for prediction of the dipole moment amplitude leading up to solar minimum. This is due to the fact that much more flux is being added to the model during solar maximum. With more flux being being advected the random motions of the convective cells have a pronounced effect. This suggests that predictions made during times of solar maximum are more difficult to make than predictions made near solar minimum.

\section{PREDICTION FOR CYCLE 24 REVERSAL}       

We used the flux transport model to predict the Solar Cycle 24 axial dipole moment reversal and subsequent magnetic field build up. Solar Cycle 14 was chosen to act as a proxy for continued active region emergence (see Figure 7, bottom). The flux transport is identical to the flux transport used in the Cycle 23 prediction: using the average axisymmetric flows and the 5 different supergranule realizations. The prediction started on 3013 August 1 and ran until 2016 December 31.

\begin{figure}[ht!]  
\centerline{\includegraphics[width=\columnwidth]{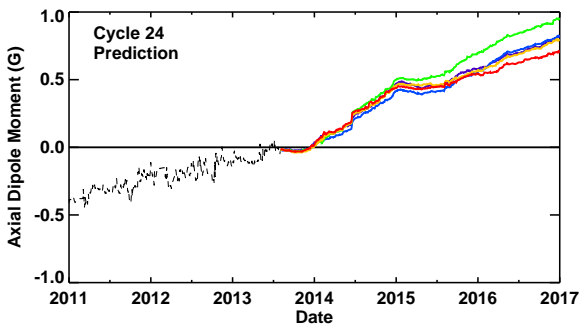}}
\caption{
Predictions of Cycle 24 with Cycle 14 active regions. Axial dipole moment predictions of Cycle 24 dipole moment reversal made with Cycle 14 active regions. For reference the observed HMI axial dipole moment is shown with a black dashed line. }
\end{figure}

The Solar Cycle 24 axial dipole moment prediction (see Figure 9) shows the dipole moment stalling for a few months before the reversal occurring in December of 2013.  The subsequent magnetic field build up is similar to the buildup observed during Solar Cycle 23. This would suggest that Solar Cycle 25 might be similar in size to Cycle 24. However, it is the axial dipole moment \emph{at solar minimum} that is the best indicator of the amplitude of the coming cycle. Minimum is not expected to occur until around 2020 or 2021. Predictions made 2-3 years prior (2017-2019) will provide a more accurate estimate of the amplitude of Cycle 25.

\section{CONCLUSIONS}     

The Sun's surface magnetic field evolution, including the buildup of the polar fields and subsequent magnetic reversals, is essential to deciphering the sunspot cycle. Previous surface flux transport models have been used to investigate the role of surface flows and active region emergence  in the surface magnetic field evolution, but all have used contrived meridional flow profiles and have parameterized the advection by supergranules as Fickian diffusion. Here we have presented a new flux transport model in which the advection by supergranules is done explicitly (rather that by parameterization). We used it to investigate metrics for defining the polar field reversals and to test the predictions of the polar fields at different phases of the solar cycle. 

We found that the axial dipole moment was the best indicator of the reversal of the Sun's magnetic field. Determinations of the polar field reversals varied by as much as a year when using different definitions of polar area.  Though it does not capture asymmetries in the polar fields, the axial dipole moment is neither ambiguous nor arbitrary. More importantly, the axial dipole moment reversal more closely reflects the timing of solar maximum and is critical for the propagation of galactic cosmic rays in the inner solar system. In the case of the synchronized solar cycle 23 polar field reversals, the axial dipole moment occurs before both the North and South reversals. Solar cycle 24 is currently experiencing an extreme asymmetry in the polar field reversal: the North has already reversed and the South is not expected to reverse until 2014. In this case the timing of the axial dipole moment reversal occurs between the North and South reversals.

We used the predictive flux transport model with Solar Cycle 23 active regions to simulate the evolution of the polar fields during the 3 years leading up to the Solar Cycle 23/24 minimum. We found that the flux transport model was able to reproduce the observed axial dipole moment without the annual signal created by presumed instrumental effect. Our supergranule flows introduce stochastic variations in the flux transport that are not captured by a diffusivity term. During this time period the stochastic nature of supergranular motions created minimal variations. 

We then used the predictive flux transport model with Solar Cycle 17 active regions to examine the predictive 
ability of the model for two different phases of the solar cycle. The first two years of results using Solar Cycle 17 for the 3 years leading up to the Solar Cycle 23/24 minimum were nearly identical to the results that used the Solar Cycle 23 active regions. While the results for the last year were somewhat divergent, they were still consistent with the baseline results. Further testing should be done to illustrate the impact of using different active region sources or a varying meridional flow. We found that our flux transport model was able to reproduce the timing of the polar field reversal of Solar Cycle 23 to within a few months at least 3 years in advance. It was shown that the stochastic nature of supergranular motions had a larger effect during this phase of the cycle. 

Results for the Solar Cycle 24 predictions show a reversal of the axial dipole moment in 2013 December. After the reversal, the axial dipole moment exhibits a rise similar in slope to the rise that followed the Cycle 23 axial dipole reversal. While this may be an early indication that Cycle 25 will be similar to Cycle 24, predictions made within 2-3 years of the coming minimum (estimated to be 2020 or 2021) will provide a more accurate estimate of the amplitude of Cycle 25.

\acknowledgements
The authors were supported by a grant from the NASA Living with a Star Program to Marshall Space Flight Center. The HMI data used are courtesy of the NASA/SDO and the HMI science team. The SOHO/MDI project was supported by NASA grant NAG5-10483 to Stanford University. SOHO is a project of international cooperation between ESA and NASA. 

\bibliography{MyBib}

\begin{thebibliography}{32}
\expandafter\ifx\csname natexlab\endcsname\relax\def\natexlab#1{#1}\fi

\bibitem[{{Babcock}(1959)}]{Babcock59}
{Babcock}, H.~D. 1959, \apj, 130, 364

\bibitem[{{Babcock}(1961)}]{Babcock61}
{Babcock}, H.~W. 1961, \apj, 133, 572

\bibitem[{{Basu} \& {Antia}(2003)}]{BasuAntia03}
{Basu}, S. \& {Antia}, H.~M. 2003, \apj, 585, 553

\bibitem[{{Basu} \& {Antia}(2010)}]{BasuAntia10}
---. 2010, \ApJ, 717, 488

\bibitem[{{de Toma}(2011)}]{deToma2011}
{de Toma}, G. 2011, \solphys, 274, 195

\bibitem[{{DeVore} {et~al.}(1984){DeVore}, {Boris}, \&
  {Sheeley}}]{DeVore_etal84}
{DeVore}, C.~R., {Boris}, J.~P., \& {Sheeley}, Jr., N.~R. 1984, \solphys, 92, 1

\bibitem[{{Dikpati} {et~al.}(2006){Dikpati}, {de Toma}, \&
  {Gilman}}]{Dikpati_etal06}
{Dikpati}, M., {de Toma}, G., \& {Gilman}, P.~A. 2006, \grl, 33, 5102

\bibitem[{{Ferreira} \& {Potgieter}(2004)}]{FerreiraPotgieter04}
{Ferreira}, S.~E.~S. \& {Potgieter}, M.~S. 2004, \apj, 603, 744

\bibitem[{{Gonz{\'a}lez Hern{\'a}ndez} {et~al.}(2010){Gonz{\'a}lez
  Hern{\'a}ndez}, {Howe}, {Komm}, \& {Hill}}]{GonzalezHernandez_etal10}
{Gonz{\'a}lez Hern{\'a}ndez}, I., {Howe}, R., {Komm}, R., \& {Hill}, F. 2010,
  \ApJ, 713, L16

\bibitem[{{Hale} {et~al.}(1919){Hale}, {Ellerman}, {Nicholson}, \&
  {Joy}}]{Hale_eta19}
{Hale}, G.~E., {Ellerman}, F., {Nicholson}, S.~B., \& {Joy}, A.~H. 1919, \apj,
  49, 153

\bibitem[{{Hathaway}(2010)}]{Hathaway10}
{Hathaway}, D.~H. 2010, Living Reviews in Solar Physics, 7, 1

\bibitem[{{Hathaway} \& {Rightmire}(2010)}]{HathawayRightmire10}
{Hathaway}, D.~H. \& {Rightmire}, L. 2010, Science, 327, 1350

\bibitem[{{Hathaway} \& {Rightmire}(2011)}]{HathawayRightmire11}
---. 2011, \ApJ, 729, 80

\bibitem[{{Hathaway} {et~al.}(2010){Hathaway}, {Williams}, {Dela Rosa}, \&
  {Cuntz}}]{Hathaway_etal10}
{Hathaway}, D.~H., {Williams}, P.~E., {Dela Rosa}, K., \& {Cuntz}, M. 2010,
  \apj, 725, 1082

\bibitem[{{Howard}(1991)}]{Howard1991}
{Howard}, R.~F. 1991, \solphys, 132, 49

\bibitem[{{Jin} {et~al.}(2013){Jin}, {Harvey}, \& {Pietarila}}]{Jin_etal12}
{Jin}, C.~L., {Harvey}, J.~W., \& {Pietarila}, A. 2013, \apj, 765, 79

\bibitem[{{Mu{\~n}oz-Jaramillo} {et~al.}(2013){Mu{\~n}oz-Jaramillo},
  {Balmaceda}, \& {DeLuca}}]{MuozJaramillo_etal13}
{Mu{\~n}oz-Jaramillo}, A., {Balmaceda}, L.~A., \& {DeLuca}, E.~E. 2013,
  Physical Review Letters, 111, 041106

\bibitem[{{Mu{\~n}oz-Jaramillo} {et~al.}(2012){Mu{\~n}oz-Jaramillo}, {Sheeley},
  {Zhang}, \& {DeLuca}}]{MuozJaramillo_etal12}
{Mu{\~n}oz-Jaramillo}, A., {Sheeley}, N.~R., {Zhang}, J., \& {DeLuca}, E.~E.
  2012, \apj, 753, 146

\bibitem[{{Norton} \& {Gallagher}(2010)}]{NortonGallagher10}
{Norton}, A.~A. \& {Gallagher}, J.~C. 2010, \solphys, 261, 193

\bibitem[{{Rightmire-Upton} {et~al.}(2012){Rightmire-Upton}, {Hathaway}, \&
  {Kosak}}]{RightmireUpton_etal12}
{Rightmire-Upton}, L., {Hathaway}, D.~H., \& {Kosak}, K. 2012, \ApJ, 761, L14

\bibitem[{{Roudier} {et~al.}(2009){Roudier}, {Rieutord}, {Brito}, {Rincon},
  {Malherbe}, {Meunier}, {Berger}, \& {Frank}}]{Roudier_etal09}
{Roudier}, T., {Rieutord}, M., {Brito}, D., {Rincon}, F., {Malherbe}, J.~M.,
  {Meunier}, N., {Berger}, T., \& {Frank}, Z. 2009, \aap, 495, 945

\bibitem[{{Scherrer} {et~al.}(1995){Scherrer}, {Bogart}, {Bush}, {Hoeksema},
  {Kosovichev}, {Schou}, {Rosenberg}, {Springer}, {Tarbell}, {Title},
  {Wolfson}, {Zayer}, \& {MDI Engineering Team}}]{Scherrer_etal95}
{Scherrer}, P.~H., {Bogart}, R.~S., {Bush}, R.~I., {Hoeksema}, J.~T.,
  {Kosovichev}, A.~G., {Schou}, J., {Rosenberg}, W., {Springer}, L., {Tarbell},
  T.~D., {Title}, A., {Wolfson}, C.~J., {Zayer}, I., \& {MDI Engineering Team}.
  1995, \solphys, 162, 129

\bibitem[{{Scherrer} {et~al.}(2012){Scherrer}, {Schou}, {Bush}, {Kosovichev},
  {Bogart}, {Hoeksema}, {Liu}, {Duvall}, {Zhao}, {Title}, {Schrijver},
  {Tarbell}, \& {Tomczyk}}]{Scherrer_etal12}
{Scherrer}, P.~H., {Schou}, J., {Bush}, R.~I., {Kosovichev}, A.~G., {Bogart},
  R.~S., {Hoeksema}, J.~T., {Liu}, Y., {Duvall}, T.~L., {Zhao}, J., {Title},
  A.~M., {Schrijver}, C.~J., {Tarbell}, T.~D., \& {Tomczyk}, S. 2012, \solphys,
  275, 207

\bibitem[{{Schrijver} \& {Title}(2001)}]{SchrijverTitle01}
{Schrijver}, C.~J. \& {Title}, A.~M. 2001, \ApJ, 551, 1099

\bibitem[{{Shiota} {et~al.}(2012){Shiota}, {Tsuneta}, {Shimojo}, {Sako},
  {Orozco Su{\'a}rez}, \& {Ishikawa}}]{Shiota_etal12}
{Shiota}, D., {Tsuneta}, S., {Shimojo}, M., {Sako}, N., {Orozco Su{\'a}rez},
  D., \& {Ishikawa}, R. 2012, \ApJ, 753, 157

\bibitem[{{Simon} {et~al.}(1988){Simon}, {Title}, {Topka}, {Tarbell}, {Shine},
  {Ferguson}, {Zirin}, \& {SOUP Team}}]{Simon_etal88}
{Simon}, G.~W., {Title}, A.~M., {Topka}, K.~P., {Tarbell}, T.~D., {Shine},
  R.~A., {Ferguson}, S.~H., {Zirin}, H., \& {SOUP Team}. 1988, \apj, 327, 964

\bibitem[{{Svalgaard} {et~al.}(2005){Svalgaard}, {Cliver}, \&
  {Kamide}}]{Svalgaard_etal05}
{Svalgaard}, L., {Cliver}, E.~W., \& {Kamide}, Y. 2005, \grl, 32, 1104

\bibitem[{{Svalgaard} \& {Kamide}(2013)}]{SvalgaardKamide13}
{Svalgaard}, L. \& {Kamide}, Y. 2013, \apj, 763, 23

\bibitem[{{Ulrich} \& {Tran}(2013)}]{UlrichTran13}
{Ulrich}, R.~K. \& {Tran}, T. 2013, \apj, 768, 189

\bibitem[{{van Ballegooijen} {et~al.}(1998){van Ballegooijen}, {Cartledge}, \&
  {Priest}}]{vanBallegooijen_etal98}
{van Ballegooijen}, A.~A., {Cartledge}, N.~P., \& {Priest}, E.~R. 1998, \ApJ,
  501, 866

\bibitem[{{V{\"o}gler} {et~al.}(2005){V{\"o}gler}, {Shelyag}, {Sch{\"u}ssler},
  {Cattaneo}, {Emonet}, \& {Linde}}]{Vogler_etal05}
{V{\"o}gler}, A., {Shelyag}, S., {Sch{\"u}ssler}, M., {Cattaneo}, F., {Emonet},
  T., \& {Linde}, T. 2005, \aap, 429, 335

\bibitem[{{Wang} {et~al.}(1989){Wang}, {Nash}, \& {Sheeley}}]{Wang_etal89}
{Wang}, Y.-M., {Nash}, A.~G., \& {Sheeley}, Jr., N.~R. 1989, \apj, 347, 529

\end{thebibliography}
\bibliographystyle{apj}

\end{document}